\documentclass[aip,pop,reprint,showpacs,preprintnumbers,amsmath,amssymb,a4paper,numerical]{revtex4-1}

\usepackage{graphicx}

\begin{document}

\preprint{Accepted Manuscript}

\title{Size and Density Evolution of a Single Microparticle Embedded in a Plasma}


\author{Oguz Han Asnaz}
\email[]{asnaz@physik.uni-kiel.de}
\affiliation{IEAP, Christian-Albrechts-Universit\"at, 24098 Kiel, Germany}
\author{Hendrik Jung}
\affiliation{IEAP, Christian-Albrechts-Universit\"at, 24098 Kiel, Germany}
\author{Franko Greiner}
\affiliation{IEAP, Christian-Albrechts-Universit\"at, 24098 Kiel, Germany}
\author{Alexander Piel}
\affiliation{IEAP, Christian-Albrechts-Universit\"at, 24098 Kiel, Germany}


\date{\today}

\hyphenation{na-no-to-po-lo-gy}
\begin{abstract} 
This article presents two measurement techniques to determine the diameter of a single dust particle during plasma operation.
Using long-distance microscopy (LDM), the particle is imaged from outside the plasma chamber.
In combination with phase-resolved resonance measurements the development of the volume-averaged particle mass density is measured over several hours.
The measurements show a significant decrease of mass density for polymethyl methacrylate (PMMA) particles due to a plasma etching process on the surface.
This is explained by a core-shell model and is supported by a surface roughness effect seen in the LDM images, an out-of-focus imaging of the angular Mie scattering pattern and ex-situ laser scattering microscopy measurements.

\end{abstract}


\maketitle

\section{Introduction}\label{sec:intro} 
In recent years ever more precise methods of measuring the plasma-dust and dust-dust interaction were developed.
One goal of these methods is to use dust particles as small, reliable probes for determining different plasma properties. \cite{Maurer11a}
The strength of the interaction forces in these systems scales differently with the particle dimensions and -- for spherical particles -- can be described by the dependency on the particle radius $a_\mathrm{p}$.

The charging of the particle described by the OML model. \cite{Langmuir26a} leads to electric field forces scaling linearly with $a_\mathrm{p}$
Drag forces, induced by the neutral gas atoms or ions scale with $a_\mathrm{p}^2$.
Finally, the gravitational force depends on the particle mass and is therefore proportional to $a_\mathrm{p}^3$.

Because of this range of dependencies most methods analyzing the trajectory of the particle need accurate knowledge about its dimensions.
When analyzing plasma inherent processes like stochastic heating through charge fluctuations \cite{Schmidt15a,Schmidt16a}, the dust radius is important for interpreting the charging mechanism.
Other experiments, like the resonance method \cite{Melzer94a,Trottenberg95a,Carstensen11a,Carstensen13a}, investigate the resonant response of the particle to small external forces.
The resulting damped oscillation of the particle enables us to determine the coefficient of friction with the neutral gas as well as the particle charge with high accuracy.
For a detailed description of the method and its applications see Ref.\,\onlinecite{Jung16a}.

With a general trend in the dusty plasma community away from phenomenological descriptions to precise measurements and modeling, knowledge about the particle radius, given by the manufacturer with an accuracy of about $\pm5\%$, became insufficient. 
When higher precision is needed, one way is to collect the particle afterwards for an ex-situ measurement, e.g., with a scanning electron microscope. \cite{Karasev16a}
For our case, this approach poses a great engineering problem to ensure that the collected particle is the one that has been used in the experiment.
Furthermore, similar measurements reveal a complex restructuring of the particle surface after prolonged plasma exposure. \cite{Coen03a}
To be able to use a single particle for a long-duration measurement, these effects need to be examined and modelled for different exposure durations by a series of ex-situ measurements.
Therefore, measurements of the particle size during plasma operation are preferable.

This article aims to explore the possibilities of optical measurements of the particle size by means of long-distance microscopy (LDM).
This measurement is combined with the resonance method to fully characterize the temporal evolution of the particle's radius and mass density.
The results are supported by ex-situ laser scanning microscopy (LSM) images of a particle after plasma exposure.

\section{Experimental Setup} 
The experiments were performed in an asymmetric capacitively coupled radio frequency (rf) discharge with an argon gas pressure of $p = (9.90\pm0.05)\,\mathrm{Pa}$ and an rf power of $2\,\mathrm{W}$ at $13.56\,\mathrm{MHz}$.
For horizontal confinement of the particle, a cylindrical depression with $20\,\mathrm{mm}$ diameter and $2\,\mathrm{mm}$ depth was milled out of the lower electrode.
The dust particles are introduced into the chamber with a salt shaker type dust dropper.

\begin{figure}
\includegraphics{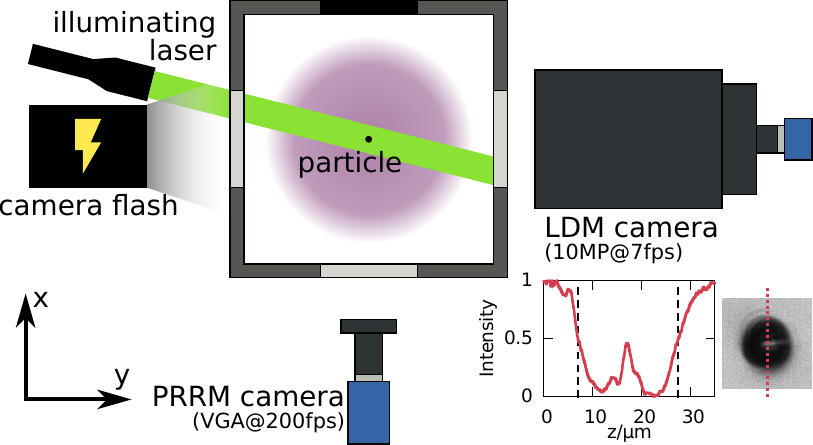}%
\caption{Top-view of the experimental setup for the combined PRRM + LDM measurement. A particle is confined in the center of the plasma chamber. It is illuminated either by a camera flash or an expanded laser and imaged by the LDM or PRRM camera respectively. The particle size is measured by the FWHM of the intensity in the LDM image.}
\label{fig:PRRMLDM}
\end{figure}

While a traditional imaging of the particle can be accomplished easily ex-situ with high precision using various types of microscopes, measurements during plasma operation are restricted by the distance from an access window to the particle.
Therefore, a \textit{Questar QM100} long-distance microscope (LDM) with a minimal working distance of about $12\,\mathrm{cm}$ has been used (see Fig.\,\ref{fig:PRRMLDM}).
The particle is illuminated by a camera flash positioned opposite to the camera.
As an incoherent light source it avoids diffraction fringes, which would obscure the contour of the particle.
It provides a short light pulse of high-intensity, which allows using a CMOS camera with a small pixel size of $1.67\,\mu\mathrm{m} \times 1.67\,\mu\mathrm{m}$.
Most cameras with such small pixels only support a rolling shutter to maximize their fill factor.
By operating the camera flash at its shortest flash duration of $40\,\mu\mathrm{s}$ and choosing a comparatively long exposure time, it is possible to avoid conflicts with the rolling shutter and motion blur due to the stochastic motion of the particle.

The point spread function (PSF) of the LDM is highly anisotropic due to the $60\,\mathrm{mm} \times 120\,\mathrm{mm}$ sized access windows, the aperture of the LDM, and the position of the LDM in relation to the access window.
The diffraction limit of our setup can be approximated as $\Delta x \approx 4\,\mathrm{\mu m}$ and $\Delta z \approx 2\,\mathrm{\mu m}$.
To acquire the particle radius with a high precision, the average FWHM of several profiles through the center of the particle is taken with the error given by its standard deviation.
Our measurements have shown an average sizing error of $\overline{\Delta a_\mathrm{p}} = 0.25\,\mathrm{\mu m}$.
The systematic error of this process is small as long as the radius of the particle is larger than the FWHM of the PSF in each direction, since the FWHM operation is invariant against a convolution with the PSF.
If this condition is not fulfilled, the particles becomes more and more indistinguishable from a point source and the particle size can therefore not be assessed correctly.

In addition to the LDM imaging, a measurement with the phase-resolved resonance method (PRRM) is performed to determine the particle mass density.
This is accomplished by biasing the lower electrode with a low-frequency sinusoidal signal, which for small amplitudes drives a damped harmonic oscillation of the particle in $z$-direction:
\begin{equation}
    \xi(\phi) = z(\phi)-z_0 = A\exp(i\phi)\,,
\end{equation}
where $z_0$ describes the resting position of the particle and $A$ the complex amplitude of the oscillation.
By measuring the height over the lower electrode at phases of $\phi = 0^\circ$, $90^\circ$, $180^\circ$, and $270^\circ$, it is possible to determine the complex amplitude. \cite{Jung16a}
A sweep over different driving frequencies and a subsequent fit to the theoretical model leads to the eigenfrequency $\omega_0$ and the gas friction coefficient $\gamma$.
The volume-averaged mass density $\bar\rho_\mathrm{p}$ of a spherical particle can then be calculated from $\gamma$ and the measured particle radius $a_\mathrm{p}$ using the Epstein formula: \cite{Epstein24a}
\begin{equation}
    \bar\rho_\mathrm{p} = \delta \frac{4}{\pi}\frac{p}{a_\mathrm{p} \gamma \bar v_\mathrm{th,n}}\,.
\end{equation}
The thermal velocity $v_\mathrm{th,n}$ of the neutral gas atoms depends on the gas type and temperature.
The gas pressure $p$ is measured with a precision capacitive pressure gauge during the experiment.
Plastic particles in rf argon plasmas were found to be described by $\delta = 1.44$. \cite{Jung15a}

For the PRRM measurements a $10\,\mathrm{mW}$ laser with an expanded cross-section is used for continuous illumination of the particle.
The particle position is measured by a CCD camera at VGA resolution with a maximum frame rate of $200\,\mathrm{fps}$.
A LabVIEW-based data acquisition software alternates between PRRM and LDM measurements and makes it possible to measure a single particle continuously over several hours.

\section{Results}\label{sec:results} 
\begin{figure}
\includegraphics{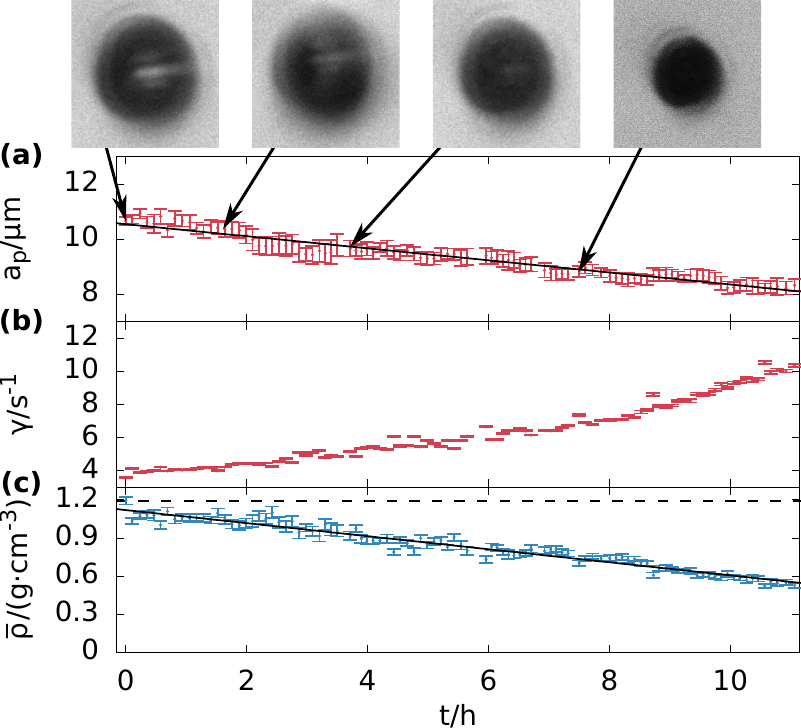}%
\caption{Long-duration measurement of a PMMA particle over 11 hours. \textbf{(a)} The radius was measured using LDM. Four images are shown exemplarily. \textbf{(b)} The coefficient of friction was determined with the PRRM. \textbf{(c)} A combination of both measurements provides the mass density of the particle.\label{fig:massdensity}}
\end{figure}

A spherical polymethyl methacrylate (PMMA) particle was confined in the plasma sheath above the lower electrode.
Using the previously described setup for combined PRRM + LDM imaging the particle was observed for over 11 hours.
Fig.\,\ref{fig:massdensity} shows the temporal development of the particle radius, friction coefficient and mass density.
The particle starts with a radius of $a_\mathrm{p} = (10.67\pm0.15)\,\mu\mathrm{m}$ and a mass density of $\bar\rho_\mathrm{p} = (1.19 \pm 0.04)\,\mathrm{g}/\mathrm{cm}^3$, which is well within the manufacturer's specification of $a_\mathrm{PMMA} = (10.01 \pm 0.14)\,\mu\mathrm{m}$ and $\rho_\mathrm{PMMA} = 1.19 \,\mathrm{g}/\mathrm{cm}^3$.
Due to the plasma exposure, the radius then decreases linearily with a rate of ${\mathrm{d}a_\mathrm{p}}/{\mathrm{d}t} = -0.22\,\mu\mathrm{m}/\mathrm{h}$.
Excepting the first data point, the averaged mass density also shows a linear decrease at a rate of ${\mathrm{d}\bar\rho_\mathrm{p}}/{\mathrm{d}t} = -0.05\,\mathrm{g}/(\mathrm{cm^3\cdot h})$.
The faster loss of density in the first 7 to 15 minutes of the experiment could be due to degassing of bound water.
This was commonly assumed to result in a mass loss of about 10 percent \cite{Pavlu04a,Carstensen11a}, which is in good agreement with our measurement.

While the linear loss of mass was known from similar experiments with PMMA particles in rf plasmas \cite{Carstensen13a}, it was assumed to be exclusively a loss of radius with the mass density staying constant.
Yet, here we observe a substantial loss of the average mass density in 11 hours to well under half the starting value.
While a definitive explanation of the loss of density is beyond the scope of this article, we want to present one plausible mechanism at this point.

Investigations of the effects of rf plasma exposure using different gases have shown, that even non-reactive gases as argon can severely modify the micro- and nanotopology of a PMMA foil. \cite{Coen03a}
It was shown, that the impinging ions from the sheath around the foil induce a ``kind of melting process involving chain scission and etching effects''. \cite{Coen03a}
The ions are accelerated in the plasma sheath and thereby impact on the particle surface with an energy of several eV.
Despite the fact, that the ions only have a penetration depth of the order of a few nanometers, the reformation of the PMMA surface propagates slowly into the bulk of the material and leads to the formation of $100\,\mathrm{nm}$ to $1\,\mathrm{\mu m}$ deep grooves after only 15 minutes of plasma exposure.
It is possible, that over time this effect can propagate deeper into the bulk of the observed PMMA particle and lead to the observed loss of density.

\begin{figure}
\includegraphics{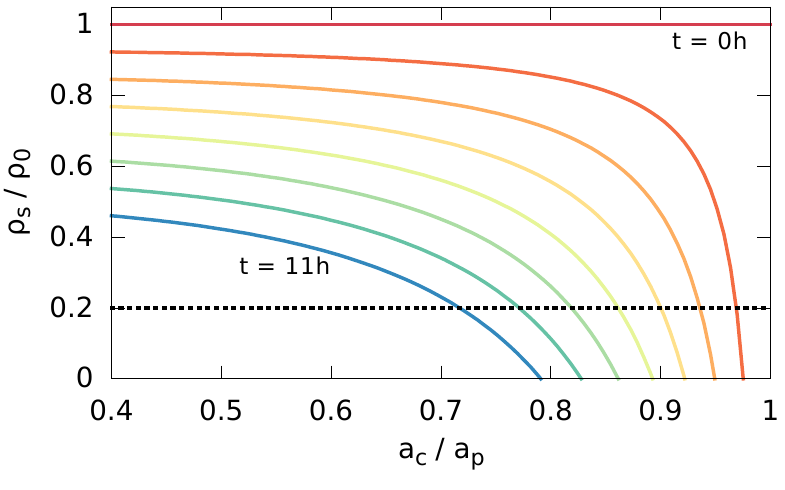}%
\caption{Results of a core-shell model of the surface etching process.
    The curves show the possible combinations of radius and density ratios for the measured average mass densities $\bar\rho$ at different times.
    A line with a constant density ratio of $0.2$ is shown as a dashed line.
}
\label{fig:surface-sim}
\end{figure}
Since an overall loss of 50 percent of mass density seems unlikely and would possibly even destabilize the particle, we suggest a core-shell model for the particle structure.
A model with an unaffected core and a more heavily etched shell can also explain the average density loss.
We model the particle with a total radius of $a_\mathrm{p}$ consisting of a core with a radius of $a_\mathrm{c}$ and a density of $\rho_0$ surrounded by a shell with a density of $\rho_\mathrm{s}$.
Comparing the masses of our volume-averaged measurement and this model leads to following equation:
\begin{eqnarray}
    \bar\rho a_\mathrm{p}^3 &=& \rho_0 a_\mathrm{c}^3 + \rho_\mathrm{s} (a_\mathrm{p}^3-a_\mathrm{c}^3) \,,\nonumber\\
    \Rightarrow \frac{\rho_\mathrm{s}}{\rho_0} &=& \frac{{\bar\rho}/{\rho_0} - \left({a_\mathrm{c}}/{a_\mathrm{p}}\right)^3}{1-\left({a_\mathrm{c}}/{a_\mathrm{p}}\right)^3}\,.
\end{eqnarray}
In Fig. \ref{fig:surface-sim} possible combinations of mass density and radii ratios are shown for average mass densities at different times of our long-term measurement.
If we assume that the etching process leads to a mass density of e.g. 20~\% in the shell region as indicated by the dashed line, this would imply that the extent of the shell gets larger over time and after 11 hours results in a shell with a thickness of 30~\% of the particle radius, which corresponds to 45~\% of the particle volume.

The surface modification hypothesis is supported by three further findings:
In LDM images at the start of the experiment, the camera flash can be seen due to a focusing effect by the spherical particle (see Fig.\,\ref{fig:massdensity}(a)) but becomes increasingly diffuse until it is fully obscured after about four hours.
This effect could result from increasingly deep grooves on the particles surface in the order of magnitude of the wavelength.

\begin{figure}
\includegraphics{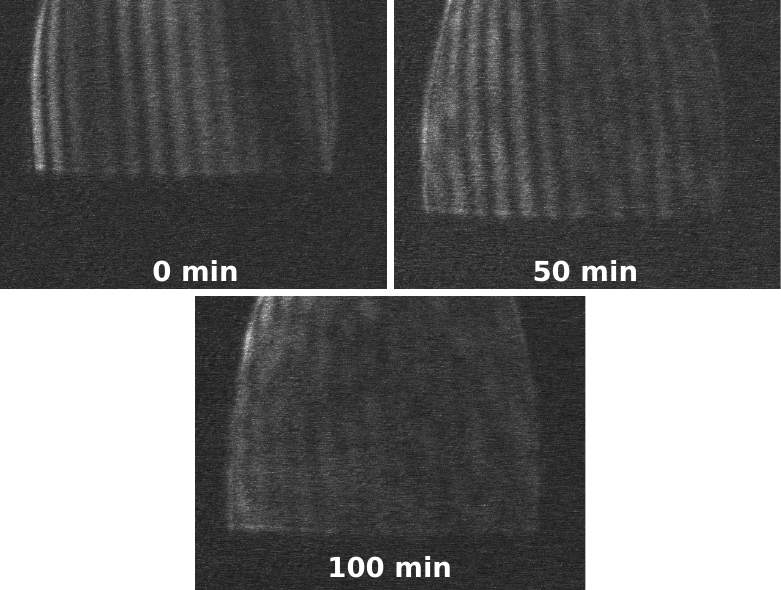}%
\caption{ILIDS images of a PMMA particle at different times. The effects of plasma exposure generate an increasing additional scattering pattern.}
\label{fig:ilids-surface}
\end{figure}
Further evidence was collected using an adapted setup for ``Interferometric Laser Imaging for Droplet Sizing'' (ILIDS) \cite{Koenig86a,Ragucci90a,Glover95a,Kawaguchi02a,Chaudhuri16a} based on the angular resolution of the Mie scattering pattern \cite{Hulst} using an out-of-focus image of the particle.
We replaced the classical glass lens with a spherical mirror with a diameter of $127\,\mathrm{mm}$ and a focal length of $101.6\,\mathrm{mm}$ to be able to achieve high numerical apertures despite of a working distance of over $10\,\mathrm{cm}$.
While our current plasma chamber setup limits the maximum imaging angle to $24.4^\circ$, a future setup with larger access windows could lead to imaging angles of up to $64^\circ$.

At first, the particle in Fig.\,\ref{fig:ilids-surface} shows a clear angular pattern, which corresponds to a particle with a size of $(9.3\pm1.8)\,\mathrm{\mu m}$ using an approximative approach described in Ref.\,\onlinecite{Glantschnig81a}.
Over time an additional pattern emerges, which becomes increasingly dominant and flickers rapidly in video recordings.
This further reinforces our hypothesis of grooves on the particle surface, which would result in an additional scattering term, which is not considered by traditional Mie theory for spherical particles.
A similar scattering phenomenon was reported for the melting of small frozen water droplets. \cite{Ruiz14a}
The flickering could indicate fast particle rotation, but did not seem to be periodic in video recordings at a frame rate of $200\,\mathrm{fps}$.

\begin{figure}
\includegraphics{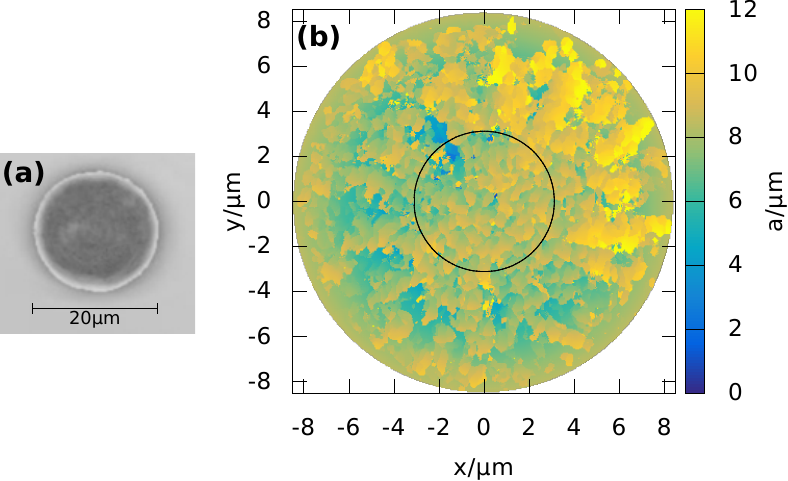}%
\caption{\textbf{(a)} Reconstructed image of an untreated PMMA particle. \textbf{(b)} Radial height map of a plasma-treated PMMA particle. Further analysis was done on the marked central region.}
\label{fig:lsm-images}
\end{figure}
\begin{figure}
\includegraphics{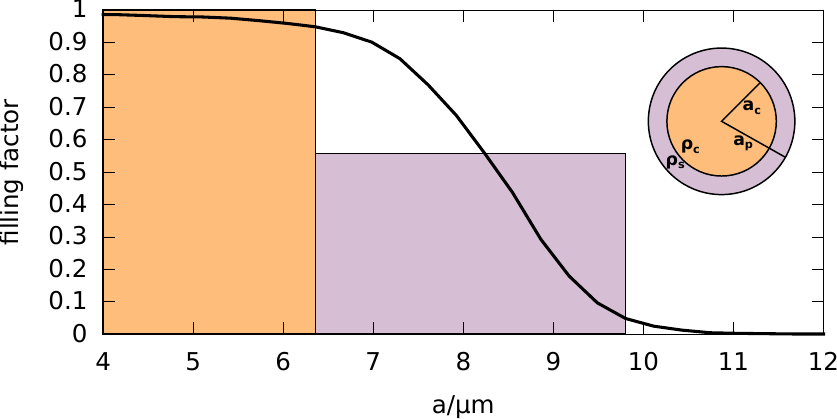}%
\caption{Volume filling factor of a particle at different radii. Black line: Calculated filling factors from analysis of the marked region in Fig.\,\ref{fig:lsm-images}. Orange: Core region at a filling factor of 1. Purple: Shell region with an average filling factor of $0.56$. Boundaries were defined with a threshold value of $0.95$ and $0.05$ respectively.}
\label{fig:lsm-profile}
\end{figure}

To qualitatively verify the core-shell model, we have performed an ex-situ measurement with another batch of PMMA microparticles.
Using a \emph{Keyence VK-9700} laser scanning microscope (LSM), the surface of a single particle was measured before and after plasma exposure (see Fig.\,\ref{fig:lsm-images}).
Since the LSM can only measure the height above the background, only measurements on the marked central circle were considered for further analysis to ensure that the height profile of the particle is not obscured by projection effects.
The radial density profile was calculated by integrating over the height distribution (see Fig.\,\ref{fig:lsm-profile}).
Applying a threshold at a filling factor of $0.95$ and $0.05$ respectively allows for a comparison between this measurement and the core-shell model and gives values of $\rho_\mathrm{s}/\rho_0 = 0.56$ and $a_\mathrm{c}/a_\mathrm{p} = 0.65$.
This confirms that the particle has lost a considerable amount of density in the shell region, while there exists a core region, which is unaffected by the plasma.

\section{Conclusion \label{sec:conclusion}} 
It has been shown that accurate measurements of the particle radius are feasible using either traditional LDM imaging or more elaborate methods based on Mie scattering.
A combination of LDM imaging and PRRM measurements on a PMMA particle has shown a temporal decrease of the particle radius with a rate of ${\mathrm{d}a_\mathrm{p}}/{\mathrm{d}t} = -0.22\,\mu\mathrm{m}/\mathrm{h}$ and of the mass density with a rate of ${\mathrm{d}\bar\rho_\mathrm{p}}/{\mathrm{d}t} = -0.05\,\mathrm{g}/(\mathrm{cm^3\cdot h})$.
This was explained in terms of a core-shell model of increasingly deeper grooves in the shell region of the particle, which result from plasma inherent etching and melting processes.
This hypothesis was supported by LDM images, an additional scattering pattern in ILIDS measurements and ex-situ LSM measurements.
The ILIDS method was demonstrated using an adapted setup with a spherical mirror instead of a glass lens, which could in future surpass the particle sizing accuracy of the LDM setup.

\begin{acknowledgements}
This work was financially supported by Deutsche Forschungsgemeinschaft (DFG) within the Transregional Collaborative Research Center SFB-TR 24, Project A2.
We thank J.~Benedikt for providing the LSM measurements.
\end{acknowledgements}

\end{document}